\documentclass[12pt]{article}
\newcommand{\ep}{\epsilon}
\newcommand{\p}{{\bf P}}
\newcommand{\CC}{{\bf C}}
\newcommand{\s}{{\bf S}}

\newcommand{\CO}{{\mathcal O}}
\newcommand{\pw}{{\cal P}}
\newcommand{\bv}{\bar{v}}
\newcommand{\bz}{\bar{z}}
\newcommand{\ap}{a'}
\newcommand{\bp}{b'}
\newcommand{\adj}{{\rm adj}}
\newcommand{\C}{{\cal C}}
\newcommand{\D}{{\cal D}}
\newcommand{\G}{{\cal G}}
\newcommand{\F}{{\cal F}}
\newcommand{\diag}{{\rm diag}}
\newcommand{\R}{{\mathbf R}}

\title{\bf Gravitational Instantons\\
\bf of  Type $\mathbf{D_k}$}
\author{
Sergey A. Cherkis\thanks{e-mail: cherkis@ias.edu}\\
\it School of Natural Sciences\\
\it Institute for Advanced Study\\
\it Einstein Drive\\
\it Princeton, NJ 08540\\
\it USA
 \rm
\and
Nigel J. Hitchin\thanks{e-mail: hitchin@maths.ox.ac.uk}\\
\it Mathematical Institute\\
\it 24-29 St Giles\\
\it Oxford OX1 3LB\\
\it UK
}

\begin{document}
\begin{titlepage}

\renewcommand{\thepage}{ }

\maketitle

\begin{abstract}
We use two different methods to obtain  Asymptotically Locally Flat
hyperk\"ahler metrics of  type $D_k$.
\end{abstract}



\end{titlepage}
\pagestyle{headings}

\section{Introduction}

We give in this paper explicit formulas for asymptotically locally flat (ALF)
hyperk\"ahler metrics  of type $D_k$. The $A_k$ case has been known for a long time as
the multi-Taub-NUT metric of  Hawking \cite{H1}, and the $D_0$ case is the 2-monopole
moduli space calculated in \cite{AH}. The $D_2$ case was highlighted  in \cite{NH1} as an
approximation to the K3 metric. Both the derivation and formulas for the $D_0$ metrics
benefited from the presence of continuous symmetry groups whereas the general $D_k$ case
has none, which perhaps explains the more complicated features of what follows. The
actual manifold on which the metric is defined is, following \cite{singular}, most
conveniently taken to be a hyperk\"ahler quotient by a circle action on the moduli space
of $U(2)$  monopoles of charge $2$ with singularities at the $k$ points
${q}_1,\dots,{q}_k\in \R^3$.

The original interest in physics of self-dual gravitational instantons (of which these
metrics are   examples)  was motivated by their appearance in the late seventies in the
formulation of Euclidean quantum gravity \cite{GH,H1,H2}. In this context they play a
role similar to that of self-dual Yang-Mills solutions in quantum gauge theories
\cite{G2}. Since then, however, these objects have appeared in various problems of
quantum gauge theory, string theory and M-theory, some of which we now mention. For
concreteness, we concentrate on  the case of the $D_k$-type ALF gravitational instanton:
\begin{itemize}
\item compactifications of supergravity, string and M-theory on self-dual gravitational
instantons preserve half  the amount of  supersymmetry of the original theory -- M-theory
on $D_k$ ALF spaces, for example, emerges as a strong string coupling limit of type IIA
string theory with an O6-plane in the presence of $k$ D6-branes \cite{Sen}
\item as
discussed by Seiberg and Witten \cite{SW3}, quantum moduli spaces of supersymmetric
${\mathcal N}=4$ $SU(2)$ gauge theories with $k$ fundamental hypermultiplets in three
dimensions are $D_k$ ALF spaces
\item these spaces can also be considered as moduli
spaces of solutions of Bogomolny equations with prescribed singularities, or as
 moduli spaces of instantons on multi-Taub-NUT spaces
that are invariant with respect to the  $S^1$ symmetry \cite{A}.
\end{itemize}
We derive the formulas by two different methods. The first is based on  unpublished work
of the second author carried out for the ALE case during a visit to the University of
Bonn in 1979. It followed the twistor approach to the $A_k$ case in
\cite{NH2}, except that one needed a polynomial solution to $x^2-zy^2=a$ (``Pell's
equation"), where $z(\zeta)$ is a quartic, instead of the simple factorization $xy=a$ as
in the $A_k$ case. This was solved by introducing the elliptic curve $w^2=z(\zeta)$ and
trying to factorize $a=(x-wy)(x+wy)$ into elliptic functions. There is a divisor class
constraint (considered in 1828 by Abel \cite{Abel}!) to doing this.

Given the more recent interpretation of the ALF solutions in terms of monopoles, the
analogous elliptic curve is naturally described as the spectral curve of the monopole, or
the equivalent Nahm data. As with all spectral curves, it is subject to a transcendental
constraint and the key problem in writing down the metric is to implement analytically
this constraint on $z(\zeta)$, cutting down the five coefficients of the quartic to give
four coordinates on the hyperk\"ahler manifold.

The second method uses the generalized Legendre transform construction of Lindstrom,
Ivanov and Ro\v cek \cite{IR}, \cite{LR} which has already been successfully used for
$k=0$ and has begun to be applied to the problem considered here by the first author in
\cite{sections}. In this case the quartic $z(\zeta)$ appears in a fundamental way and the
constraint  is expressed by a differential equation. We show how these two expressions
for the constraint coincide.

When the singularities of the monopole lie on a line through the origin,  the metric
defines an explicit resolution of the $D_k$ quotient singularity by a configuration of
holomorphic 2-spheres, intersecting according to the $D_k$ Dynkin diagram. We hope to
consider this special case in more detail elsewhere.

\section{Singular monopoles}

We shall use certain moduli spaces of singular monopoles to obtain our $D_k$ metrics, as
in \cite{sections}. There are two approaches to this: through Nahm's equations, which
give us a concrete analytical description of the objects in the space and the ${\mathcal
L}^2$ metric defined on it, and the twistor approach which we use here. The latter
describes a generic point in the moduli space via an algebraic curve, and allows us the
possibility of an explicit determination of the metric.

\subsection{The twistor approach to monopoles}

It was shown in \cite{singular}, using the technique developed in \cite{NJH} as well as results of Kronheimer \cite{Kronheimer},
that a charge $2$ $U(2)$ monopole solution to the Bogomolny equations with $k$ singularities is described by a spectral curve
$\s$ in $T\p^1$ and two sections $\pi$ and $\rho$ of holomorphic line bundles $L^{-\mu}(k)\vert_{\s}$
and $L^{\mu}(k)\vert_{\s}$ respectively.
Viewing $T\p^1$ as the total space
of $\CO(2)$ we denote by $\zeta$
an affine  coordinate on $\p^1$ and $\eta$ the standard linear coordinate in the fibre. Then $L^{\mu}(k)\rightarrow T\p^1$ denotes the line  bundle
with  transition function $\zeta^{-k}\exp(\mu\eta/\zeta)$.

There is a  real structure $\sigma$ given by $\sigma(\eta,\zeta)=(-\bar{\eta},
-1/\bar{\zeta})$. The spectral curve $\s\subset T\p^1$ for a charge $2$ monopole  is then
given by an equation
\begin{equation}
\eta^2-y \eta-z=0,
\end{equation}
where $y$ is a real section of $\CO(2)$ and $z$ is a real section of $\CO(4)$.
The reality condition for a section $x$ of $\CO(2n)$ is
$\overline{x(\sigma(\zeta))}=(-1)^n x(\zeta)/\zeta^{2n}.$ Thus in the
patch $\zeta\neq\infty$ the section $x$ is given by a  polynomial $x(\zeta)$ of degree $2n$ which
 satisfies $\overline{x(-1/\bar{\zeta})}=(-1)^n x(\zeta)/\zeta^{2n}.$

The position of each of the $k$ singularities of the monopole configuration can be described by a real section $q_i$ of $\CO(2)$,
$i=1,\ldots, k$. We denote these sections by $\p^1_{q_i}\subset T\p^1.$
\vskip .25cm
The two sections $\pi$ and $\rho$ on the spectral curve are interchanged by the
real structure and satisfy
\begin{equation} \label{u2}
\pi \rho =\prod_{i=1}^k \left(\eta-q_i\right).
\end{equation}
There is a circle action on this data given by
\begin{equation} \label{circle}
(\rho,\pi)\mapsto (\lambda \rho,\lambda^{-1}\pi)
\end{equation}

Equation (\ref{u2}) says that the intersection of $\s$ with all the curves $\p^1_{ q_i}$ defines (if $\s$ does not contain one of them as a component) a divisor on $\s$ of degree $4k$, and the constraint on the spectral curve  is expressed by the fact that we can divide these into two sets of $2k$ points, one of which is a divisor for $L^{-\mu}(k)\vert_{\s}$
and the other for $L^{\mu}(k)\vert_{\s}$.

\subsection{The twistor approach to the moduli space}

The spectral curve description fits into the twistor description of the hyperk\"ahler metric on the moduli space, which we recall from \cite{singular}, and is similar to the case of non-singular monopoles in \cite{AH}. Let
$$D=\{(\eta,y,z)\in {\CO(2)}\oplus {\CO(2)}\oplus {\CO(4)}\vert \eta^2-y\eta-z=0\}$$
which has a projection $p_1(\eta,y,z)=\eta$ onto $T\p^1$ and another  $p_2(\eta,y,z)=(y,z)$ which represents $D$ as a ramified double covering of ${\CO(2)}\oplus {\CO(4)}$. We let $V^{\mu}$ be the rank 2 holomorphic vector bundle on ${\CO(2)}\oplus {\CO(4)}$ which is the direct image sheaf
$$V^{\mu}=(p_2)_*(p_1^*L^{\mu}).$$
Then the direct image of the equation (\ref{u2}) defines a subvariety $Z$ of
$V^{\mu}(k)\oplus V^{-\mu}(k)$ which will be a model of our twistor space. There is some resolution of singularities to be carried out, but that doesn't affect the determination of the metric. In the twistor space approach we need to find the twistor lines, which are sections of $p:Z\rightarrow \p^1$. The quadratic and quartic $y,z$ define a section of ${\CO(2)}\oplus {\CO(4)}$ and the functorial property of the direct image says that $\pi,\rho$ define a lifting to $Z\subset V^{\mu}(k)\oplus V^{-\mu}(k)$.
\vskip .25cm
Over $\zeta\ne 0$, the direct image equation for $Z$ can be written as
$$(x_1+\eta x_2)(y_1+\eta y_2)=\prod_i(\eta-q_i)\,\,\,{\mathrm {mod}}\,\,  \eta^2-y\eta-z=0$$
or equivalently,
\begin{eqnarray}
x_1y_1+zx_2y_2&=&p(y,z)\nonumber\\
x_2y_1+x_1y_2+yx_2y_2&=&q(y,z)\label {twistor}
\end{eqnarray}
where
$$\prod_i(\eta-q_i)=p+\eta q\,\,\,{\mathrm {mod}}\,\,  \eta^2-y\eta-z=0.$$
This equation defines a $5$-dimensional twistor space for an $8$-dimensional hyperk\"ahler manifold -- the moduli space of charge $2$ singular monopoles. There is a symplectic form along the fibres which can be written as in \cite{sections} as
\begin{equation}\label{omega}
\omega=4\sum_{j=1}^2 \frac{d\rho(\beta_j)\wedge d\beta_j}{\rho(\beta_j)},
\end{equation}
with $\eta=\beta_j$ being roots of $\eta^2-y\eta-z=0$, but in coordinates $x_1,x_2,y,z$ above as
$$\omega=4\frac{(x_1dx_1+(yx_1-zx_2)dx_2)\wedge dy+(x_1dx_2-x_2dx_1)\wedge dz}{x_1^2-zx_2^2+yx_1x_2}$$

\subsection{The hyperk\"ahler quotient}
To obtain a $4$-dimensional manifold we shall take a hyperk\"ahler quotient by a circle action which at the twistor space level is given by (\ref{circle}) and in the above coordinates is
 $$(x_1,x_2,y_1,y_2,y,z)\mapsto (\lambda x_1,\lambda x_2,\lambda^{-1}y_1,\lambda^{-1}y_2,y,z).$$
In coordinates $x_1,x_2,y,z$ the vector field generated by this action is
$$X=x_1\frac{\partial}{\partial x_1}+x_2\frac{\partial}{\partial x_2}$$
so that
$$i(X)\omega=4dy.$$
\noindent{\bf Remark:} The moment map for this action is $4y$. From the point of view of
monopoles this  can be interpreted in terms of the centre of mass, in which case the more
natural value would be $y/2$ which amounts to a rescaling of $\omega$.
\vskip.25cm

 The hyperk\"ahler quotient from the twistor point of view is just
the fibrewise symplectic quotient, so we set the moment map $y=0$ (this means that the
centre of mass of the monopole is at the origin) and take the quotient by the ${\mathbf
C}^*$ action. Putting $y=0$ in (\ref{twistor}), gives
\begin{eqnarray}
x_1y_1+zx_2y_2&=&p(y,z)\nonumber\\
x_2y_1+x_1y_2&=&q(y,z)\label{twistor0}
\end{eqnarray}
We need a smooth $4$-manifold so the circle action on the zero set of the hyperk\"ahler moment map must be
 free. In the nonsingular monopole case this is  automatic: up to a finite covering the moduli space
 is isometric to a product $M_k^0\times S^1\times {\mathbf R}^3$. The action is not necessarily free
  in the singular case.  For example, the moduli space of charge $1$ monopoles  with $k$
  singularities is the $A_{k-1}$ ALF space -- multi-Taub-NUT space -- and the triholomorphic
  circle action has fixed points.

Suppose we have a fixed point, then there is a fixed twistor line, so on each fibre of the twistor space a fixed point, which from (\ref{circle}) is where $x_1=x_2=0$. This means $p=q=0$. But with $y=0$, $p$ and $q$ are defined by
$$r(\eta)=\prod_i(\eta-q_i)=p+\eta q\,\,\,{\mathrm {mod}}\,\,  \eta^2-z=0$$
thus
$$p(z)=\frac{1}{2}(r(\eta)+r(-\eta)),\quad q(z)=\frac{1}{2\eta}(r(\eta)-r(-\eta))$$
and $p$ and $q$ have a common zero if $r(\eta)$ and $r(-\eta)$ have a common zero, i.e.
$q_i=-q_j$ (for all $\zeta$). If $q_i=0$, then $q(0)=r'(0)$ so since the $q_i$ are
distinct, the action doesn't have a fixed point. Thus, so long as the positions of the
singularities of the monopole satisfy $q_i\ne -q_j$ for any $i\neq j$, we shall  produce
a non-singular $4$-manifold as a hyperk\"ahler quotient. \vskip .25cm The equation of the
twistor space of the quotient can be obtained by using the ${\mathbf C}^*$-invariant
coordinates
$$P=2x_1y_1-p,\quad Q=2x_2y_2-q$$
and then from (\ref{twistor0})
$$\frac{P-p}{Q-q}=z\frac{x_2}{x_1}=z\frac{Q+q}{P+p}$$ which gives
\begin{equation}
P^2-zQ^2=\prod_{i=1}^k\left(z-q_i^2\right).
\label{PQ}
\end{equation}
The holomorphic symplectic form along the fibres is then
\begin{equation}\label{symplectic}
\omega=d\left(\frac{1}{\sqrt{z}}\log\frac{P+\sqrt{z}Q}{P-\sqrt{z}Q}\right)\wedge d z.
\end{equation}
\vskip .25cm
\noindent{\bf Remark:}
Equation (\ref{PQ}) defines  a subvariety of the rank $2$ direct image vector bundle $V^{2\mu}(2k)$ over $\CO(4)$. Concretely, $V^{2\mu}$ has transition matrix $$\pmatrix{\tilde P\cr
            \tilde Q}=\pmatrix{\cosh (2\mu \sqrt{z}/\zeta) & \sqrt{z}\sinh (2\mu \sqrt{z}/\zeta)\cr
\zeta^2\sinh (2\mu \sqrt{z}/\zeta)/\sqrt{z} & \zeta^2\cosh (2\mu \sqrt{z}/\zeta )}\pmatrix{P\cr
                                      Q}.$$
Note that if $\tilde z=z/\zeta^4$ then
$$\tilde P^2-\tilde z\tilde Q^2=P^2-zQ^2$$
which describes a global holomorphic quadratic form $(u,u)$ on $V$, singular on the zero section $z=0$. The null-directions are $P\pm\sqrt{z}Q$ which are globally defined on the double covering $\CO(2)$ and are the line bundles $L^{\pm 2\mu}$.
This quadratic form gives a more invariant way of writing (\ref{PQ}):
\begin{equation}
(u, u)=\prod_{i=1}^k\left(z-q_i^2\right).
\label{Zequation}
\end{equation}
\subsection{The $D_k$ singularity}
The fibre over $\zeta \in \p^1$ in the $3$-dimensional variety defined by (\ref{Zequation}) has the equation
\begin{equation}
P^2-zQ^2=\prod_{i=1}^k\left(z+p_i^2\right)
\label{PQp}
\end{equation}
if we set $p_i=iq_i(\zeta)$. The universal deformation of the $D_k$ singularity has the form
$$x^2-zy^2=\frac{1}{z}\left(\prod_{i=1}^k(z+p_i^2)-\prod_{i=1}^kp_i^2\right)+2i\prod_{i=1}^kp_iy$$
and putting
$$P=iyz-\prod_{i=1}^kp_i,\quad Q=ix$$
we obtain (\ref{PQp}). If  the $p_i^2$ are distinct for a fixed $\zeta$ and if none of the $p_i$ vanish, (\ref{PQp}) defines a smooth surface  and  $(x,y,z)\mapsto (P,Q,z)$ is a biholomorphic map. The actual twistor space $Z$ for the hyperk\"ahler metric involves a resolution of these singularities but, as in \cite{NH2},  we only need the singular space to  compute the metric, since a generic twistor line misses the singularities.

\section{The equations defining the constraint}

The main problem in determining a metric in the twistor approach is to find the twistor lines -- holomorphic sections of the projection $Z\rightarrow\CO(4)\rightarrow \p^1$.
From the construction above, we have  a rational curve $C$ in $\CO(4)$ defined by $z=z(\zeta)$  where $z(\zeta)$ is a quartic polynomial, and to lift it further we needed a section $u(\zeta)$ of $V^{2\mu}(2k)$ over $C$, which from the functorial property of the direct image was equivalent to a section $a=P+\sqrt{z}Q$ of $L^{2\mu}(2k)$ on the spectral curve. For the section $u(\zeta)$ to satisfy (\ref{Zequation}) is equivalent to the existence of sections $a=P+\sqrt{z}Q,b=P-\sqrt{z}Q$ on the spectral curve $\s$ satisfying
$$ab=\prod_{i=1}^k\left(z-q_i^2\right).$$
The spectral curve of the monopole had sections $\pi,\rho$ such that
$$\pi \rho =\prod_{i=1}^k \left(\eta-q_i\right).$$
Setting
$$a(\eta,\zeta)=\rho(\eta,\zeta) \pi(-\eta,\zeta),\quad b(\eta,\zeta)=(-1)^k\rho(-\eta,\zeta) \pi(\eta,\zeta)$$
gives sections $a,b$  of $L^{2\mu}(2k)\vert_{\s}$ and $L^{-2\mu}(2k)\vert_{\s}$ respectively such that
$$a b=\prod_{i=1}^k\left(z-q_i^2\right)$$
is satisfied.
If the $\alpha_{ij}$ are roots of $z(\zeta)-q_i^2(\zeta)$, then  $a$ vanishes at points
$(\eta,\zeta)=\left((-1)^j q_i(\alpha_{ij}), \alpha_{ij}\right)\in\s\in T\p^1$. The section $b$
vanishes at $(\eta,\zeta)=\left(-(-1)^j q_i(\alpha_{ij}), \alpha_{ij}\right).$
\vskip .25cm
A rotation of $\R^3$  acts as a  fractional linear transformation on the coordinate
$\zeta$. Using such a transformation we can put $z(\zeta)$ in the form
$r_1\zeta^3-r_2\zeta^2-r_1\zeta$ with real $r_2$ and $r_1\geq 0$. In these coordinates we
shall solve the constraint equation for the spectral curve $\eta^2=z(\zeta)$ using
Weierstrass elliptic functions: $\eta=\sqrt{r_1}\pw'(u)/2$ and $\zeta=\pw(u)+r_2/3r_1$,
where $u$ is the affine coordinate on $\CC/\Gamma$ representing the torus
$\eta^2=z(\zeta)$. Note that $\eta\rightarrow -\eta$ corresponds to $u\rightarrow -u$.
In what follows we  order
the points $\alpha_{ij}$ so that $\rho$ vanishes at
$\alpha_{i2}$ and $\alpha_{i4},$  while $\pi$ vanishes at $\alpha_{i1}$ and $\alpha_{i3}$.
Let $u_{ij}$ be the zeros of the sections $\rho$ and $\pi$
corresponding to $\left(q_i(\alpha_{ij}),\alpha_{ij}\right)$, then the condition for the sections to be doubly periodic translates into
\begin{equation}\label{trivial}
2\mu\sqrt{r_1}+\sum_{i=1}^k\sum_{j=1}^4 (-1)^j u_{ij}=0,
\end{equation}
or in terms of $z$ and $\zeta$
\begin{equation}\label{roots}
\sum_{ij}(-1)^j\int^{\alpha_{ij}}\frac{d\zeta}{\sqrt{z}}=-4\mu.
\end{equation}

From the fact that $a/b$ is a section of $L^{4\mu}\vert_{\s}$ with zeroes at
$(-1)^j u_{ij}$ and poles at $-(-1)^j u_{ij}$ it follows  (see \cite{singular} and
\cite{sections}) that with $a=P+\sqrt{z}Q, b=P-\sqrt{z}Q$ and $\zeta_W$ the Weierstrass
zeta function
\begin{equation}
\log\frac{a}{b}=-2\mu\sqrt{r_1}\left(\zeta_W(u+u_\infty)+\zeta_W(u-u_\infty)\right)
+\log\prod_{ij}\frac{\sigma(u-(-1)^j u_{ij})}{\sigma(u+(-1)^j u_{ij})}.
\end{equation}
Returning to $z$ and $\zeta$
\begin{equation}\label{abel}
\frac{1}{\sqrt{z(\zeta)}}\log\frac{a}{b}=\sum_{ij}(-1)^j\int^{\alpha_{ij}}\frac{d\xi}{(\xi-\zeta)\sqrt{z(\xi)}}.
\end{equation}

\section{From the twistor space to the metric}\label{canonical}

The first calculation uses Penrose's original nonlinear graviton construction \cite{Pen}. First we calculate the conformal structure and then use the holomorphic form to determine the volume form. The $4$-dimensional spacetime $\mathcal M$ is the space of twistor lines, a tangent vector is a holomorphic section of the normal bundle and it is a null vector for the conformal structure if and only if that section vanishes somewhere on the twistor line. This is the infinitesimal version of the statement that two points in $\mathcal M$ are null separated if the twistor lines meet.

\subsection{The complex conformal structure}

The twistor line is given by
\begin{equation}\label{z}
\eta^2=z(\zeta)=A\prod_{j=1}^4\left(\zeta-a_j\right)
\end{equation}
and we define
\begin{equation}
\chi(\zeta)=\frac{1}{\sqrt{z}}\log\frac{a}{b}=\frac{1}{\sqrt{z}}\log\frac{P(\zeta)+\sqrt{z(\zeta)}Q(\zeta)}{P(\zeta)-\sqrt{z(\zeta)}Q(\zeta)},
\label{chi1}
\end{equation}
so that  the symplectic form given by equation (\ref{symplectic}) is
\begin{equation}
\omega=d\chi\wedge d z.
\end{equation}
We can think of the variables $a_j$ and A satisfying the constraint (\ref{roots}) as providing
four real coordinates on the $D_k$ ALF manifold $\cal M$. Later we shall change coordinates to
$z(0)$ and $Q(0).$

Varying the parameters $A$ and $a_j$ under the constraint (\ref{roots}) we are seeking the  condition for an infinitesimal variation to vanish. From now on we denote the
infinitesimal variations by a prime, e.g. $A', a_j', z'$ etc.
Putting the variation of equation (\ref{z}) to zero we have
\begin{equation}\label{zvariation}
\frac{z'}{z}=\frac{A'}{A}-\sum_{j=1}^4\frac{a'_j}{\zeta-a_j}=0,
\end{equation}
while the vanishing of the variation of the constraint (\ref{roots}) gives
\begin{equation}\label{variation}
\sum_{ij}(-1)^j\frac{\alpha'_{ij}}{\sqrt{z(\alpha_{ij})}}=\sum_{ij}(-1)^j\int^{\alpha_{ij}}\frac{d\zeta}{2\sqrt{z(\zeta)}}
\left(\frac{A'}{A}-\sum_l\frac{a'_l}{\zeta-a_l}\right),
\end{equation}
and for the second term on the right hand side we note that
\begin{eqnarray}
&&\frac{1}{2}\sum_{ij}(-1)^j\int^{\alpha_{ij}}
\frac{d\zeta}{(\zeta-a_l)\sqrt{z(\zeta)}}=\frac{1}{2}\chi(a_l)=\nonumber\\
&&=\lim_{\zeta\rightarrow a_l}\frac{1}{2\sqrt{z(\zeta)}}\log\frac{P(\zeta)+\sqrt{z(\zeta)}Q(\zeta)}{P(\zeta)-\sqrt{z(\zeta)}Q(\zeta)}=\frac{Q(a_l)}{P(a_l)},
\end{eqnarray}
as $z(a_l)=0$.
Let us introduce a function
\begin{equation}
L(\zeta)=\sum_i \frac{l_i(\zeta)}{z(\zeta)-q_i^2(\zeta)},
\end{equation}
where $l_i$ are cubic polynomials in $\zeta$ such that $l_i(\alpha_{ij})=q_i(\alpha_{ij})$ (Lagrange interpolation polynomials). Also let
\begin{equation}
l=\lim_{\zeta\rightarrow\infty}\zeta L(\zeta).
\end{equation}
Then using the constraint equation (\ref{roots}) and substituting $\alpha'_{ij}$ in equation (\ref{variation}) we find
\begin{equation}\label{constraint}
\frac{A'}{A}\left(2\mu-l\right)=\sum_{l=1}^4 a'_l\left(L(a_l)-\frac{Q(a_l)}{P(a_l)}\right)
\end{equation}
The variation of $\chi$ is
\begin{equation}
\chi'=\left(L(\zeta)-\frac{1}{2}\chi(\zeta)\right)\frac{z'}{z}+\sum_{l=1}^4 \frac{a'_l}{\zeta-a_l}\left[L(a_l)-\frac{Q(a_l)}{P(a_l)}\right].
\end{equation}
Requiring the variation of $\chi$ and $z$ to vanish gives firstly
\begin{equation}\label{chi}
\sum_{l=1}^4 \frac{a'_l}{\zeta-a_l}\left[L(a_l)-\frac{Q(a_l)}{P(a_l)}\right]=0.
\end{equation}
and secondly, eliminating $A'$ from the equations (\ref{zvariation}) and (\ref{constraint}) leads to
\begin{equation}\label{second}
\sum_{l=1}^4\frac{ a'_l}{\zeta-a_l}\left[2\mu-l+a_l\left(L(a_l)-\frac{Q(a_l)}{P(a_l)}\right)\right]=0.
\end{equation}
Equations (\ref{chi}) and (\ref{second}) give a system of two equations cubic
in $\zeta$ and linear in $a'_l$. Our task is to find the condition on the coefficients of these
equations to have a common solution  for $\zeta.$ This condition is provided by the vanishing
of the resultant $R$ of the system (\ref{chi}, \ref{second}), which is   a polynomial in $a'_l$ of
degree six. However, if $a'_j=0$ then $\zeta=a_j$ solves the equations and so   $R$ must be  divisible by $a'_1 a'_2 a'_3 a'_4$. This means
\begin{equation}
S(a'_l)=R(a'_l)/(a'_1 a'_2 a'_3 a'_4)
\end{equation}
is a polynomial which is quadratic in the variations and so  is the quadratic form defining the conformal structure.
\vskip .25cm
For convenience we introduce
\begin{equation}
C_i=L(a_i)-\frac{Q(a_i)}{P(a_i)}=L(a_i)-\frac{1}{2}\chi(a_i),\quad\ D_i=2\mu-l+a_i C_i.
\end{equation}
Computing the resultant (see Appendix) we find
\begin{equation}
S(da_i)=\left|\begin{array}{cccccc}
\sum_{i=1}^4 D_i d a_i & 0 & a_1^2D_1 &
a_2^2D_2 &a_3^2D_3 &a_4^2D_4 \\
0 & 0 & a_1 D_1 & a_2 D_2 & a_3 D_3 & a_4
D_4\\
0 & -\sum_{i=1}^4 D_i d a_i/a_i & D_1 & D_2 & D_3 & D_4\\
\sum_{i=1}^4 C_i d a_i & 0 & a_1^2C_1 &
a_2^2C_2 &a_3^2C_3 &a_4^2C_4 \\
0 & 0 & a_1 C_1 & a_2 C_2 & a_3 C_3 & a_4
C_4\\
0 & -\sum_{i=1}^4 C_i d a_i/a_i & C_1 & C_2 & C_3 & C_4\\

\end{array}\right|
\end{equation}
where  we  use differentials $da_i$ instead of the primes $a'_i.$

\subsection{The real metric}
To impose the reality condition, we note that the map $\zeta\rightarrow-1/\bar{\zeta}$ takes the set of four roots
$\{a_1, a_2, a_3, a_4\}$ into itself, so that for a permutation $\sigma:(1,2,3,4)\rightarrow(2,1,4,3)$
we have $a_{\sigma(i)}=-1/\bar{a_i}.$ Now noting that
\begin{eqnarray}
\overline{L(-\frac{1}{\bar{\zeta}})}&=&\zeta^2 L(\zeta)-l\zeta,\\
\overline{\chi(-\frac{1}{\bar{\zeta}})}&=&\zeta^2\chi(\zeta)-4\mu\zeta,
\end{eqnarray}
it follows that
\begin{eqnarray}
\overline{C_{\sigma(i)}}=a_i D_i,& &\overline{D_{\sigma(i)}}=-a_i C_i\\
\overline{a_{\sigma{(i)}}C_{\sigma(i)}}=-D_i,& &\overline{a_{\sigma(i)}D_{\sigma(i)}}=C_i\\
\overline{a^2_{\sigma{(i)}}C_{\sigma(i)}}=\frac{D_i}{a_i},& & \overline{a^2_{\sigma(i)} D_{\sigma(i)}}=-\frac{C_i}{a_i}.
\end{eqnarray}
With these relations, one can easily check that $\bar{S}=S/(a_1a_2a_3a_4),$ and the quantities
\begin{equation}
A_0=\sum_{i=1}^4 C_i d a_i,\quad\ B_0=\sum_{i=1}^4 D_i d a_i
\end{equation}
satisfy
\begin{equation}
\overline{A_0}=\sum_{i=1}^4 D_i \frac{d a_i}{a_i},\quad\ \overline{B_0}=-\sum_{i=1}^4 C_i \frac{d a_i}{a_i}.
\end{equation}
Thus, we find that the metric is proportional to the real symmetric two-form
\begin{equation}
\frac{1}{\sqrt{a_1a_2a_3a_4}}S=\left(A_0 B_0\right) G \left(\begin{array}{c}\overline{A_0}\\ \overline{B_0}\end{array}\right),
\end{equation}
with
\begin{equation}
G=\frac{1}{\sqrt{a_1a_2a_3a_4}}\left(\begin{array}{cc} [a^2 D,a D,a C,C] & -[a^2D, a D, D, a C] \\ {}[a D, a^2 C, a C, C] & [a D, D, a^2C, a C]  \end{array} \right).
\end{equation}
Here we use the notation
\begin{equation}
[e,f,g,h]=\det\left(\begin{array}{cccc}
e_1&f_1&g_1&h_1\\
e_2&f_2&g_2&h_2\\
e_3&f_3&g_3&h_3\\
e_4&f_4&g_4&h_4
\end{array}\right).
\end{equation}
Observing that
\begin{eqnarray}
a_1a_2a_3a_4&=&{[a D,D,a^2C,a C]}/{\overline{[a D,D, a^2C,a C]}}\nonumber\\
&=&-{[a^2D,a D,D,a C]}/{\overline{[a D,a^2C,a C, C]}}
\end{eqnarray}
we have $|[a D,D, a^2C,a C]|G$ given by
\begin{equation}
\left(\begin{array}{cc} [a^2D,a D, a C,C]\overline{[aD,D,a^2C,aC]} & [aD, D, a^2C,a C]\overline{[a D, a^2C, a C, C]}\\ {}
[a D, a^2C, a C, C]\overline{[aD, D, a^2C,a C]} & [a D, D, a^2C, a C]\overline{[a D, D, a^2C, a C]} \end{array} \right).
\end{equation}
Utilizing the identity relating the determinants derived in the Appendix
\begin{eqnarray}\label{relation}
& &[ a C,C,a^2 D,a D]\overline{[a^2 C,a C,a D,D]}= \nonumber\\
& &[ C,a C,a^2 C,a D]\overline{[C,a C,a^2C,a D]}+[C,a C,D,a D]^2,
\end{eqnarray}
we have
\begin{equation}\label{Gnew}
\frac{|[a D,D,a^2C,a C]|}{[a D,D,a C,C]^2}\ G=\left(\begin{array}{cc} 1+\gamma\bar{\gamma}& \bar{\gamma}\delta\\ \gamma\bar{\delta}&\delta\bar{\delta}\end{array}\right),
\end{equation}
where
\begin{equation}\label{gammadelta}
\gamma=\frac{[aD, a^2C, a C,C]}{[a D,D, a C, C]},\quad \delta=\frac{[a D,D, a^2C,a C]}{[a D,D, a C, C]}.
\end{equation}
\vskip .25cm
Next we identify the differentials $A_0$ and $B_0$ in terms of the coordinates $z(0), \chi(0)$.
\begin{equation}\label{diffs}
\overline{A_0}=(2\mu-l)\frac{d z(0)}{z(0)},\quad \overline{B_0}=d\chi(0)-\left(L(0)-\frac{1}{2}\chi(0)\right)\frac{d z(0)}{z(0)}
\end{equation}
From the definition of $\chi$ (\ref {chi1}) we have
\begin{equation}
d\chi=2\frac{dQ}{P}+\left(\frac{Q}{P}-\frac{Q}{P}\sum_{j=1}^k\frac{z}{z-q_j}-\frac{1}{2}\chi\right)\frac{d z}{z},
\end{equation}
thus
\begin{equation}
\overline{B_0}=2\frac{dQ}{P}+\left(\frac{Q}{P}-\frac{Q}{P}\sum_{j=1}^k\frac{z(0)}{z(0)-q_j}-L(0)\right)\frac{d z(0)}{z(0)}.
\end{equation}
The conformal structure, from (\ref{Gnew}) can therefore be written as
\begin{equation}
ds^2\sim A_0\overline{A_0}+(\bar{\gamma}A_0+\bar{\delta}B_0)(\gamma\overline{A_0}+\delta\overline{B_0}).
\end{equation}
The volume corresponding to the expression on the right-hand-side is
\begin{equation}
{\rm Vol}=-\frac{(2\mu-l)^2|\delta|^2}{z\bar{z}P\bar{P}}dz\wedge d\bar{z}\wedge dQ\wedge d\bar{Q}.
\end{equation}
On the other hand the volume should be equal to $\frac{1}{4}\omega\wedge\bar{\omega}$ for
$\omega$ given by equation (\ref{symplectic}). Thus comparing the two expressions for
$\omega=2 dQ\wedge d z/P$ we have the final metric
\begin{equation}\label{metric1}
ds^2=\left|\frac{z}{(2\mu-l)\delta}\right|(A_0\overline{A_0}+(\bar{\gamma}A_0+\bar{\delta}B_0)(\gamma\overline{A_0}+\delta\overline{B_0})).
\end{equation}

\section{The generalized Legendre transform}

We have obtained the hyperk\"ahler metrics above by following the twistor space approach.
In this section, we apply a different technique  given by generalized Legendre transform
\cite{LR, HKLR}. It has been successfully applied in the case of $D_0$ ALF in \cite{IR} to reproduce Atiyah-Hitchin metric, and
to study $D_k$ ALF metrics in \cite{CRW}.

We start with a brief outline of the construction. One can pick, as earlier,
$z(\zeta)\in\CO(4)$ as one of the complex coordinates on the fibres of the twistor space.
In the patch $\zeta\neq\infty$ it has the form
\begin{equation}
z(\zeta)=z+v\zeta+w\zeta^2-\bar{v}\zeta^3+\bar{z}\zeta^4.
\end{equation}
The second coordinate $\chi$ (see (\ref{chi1})) is such that the holomorphic twistorial two-form
has the form $\omega=d\chi\wedge dz$. If the section $\chi$ is represented by the function $\chi_1$
in the patch $\zeta\neq\infty$ and $\chi_2$ in the patch $\zeta\neq0$, then we
define a function $\hat{f}$ and a contour $C$ such that
\begin{equation}
\oint_C
\frac{d\zeta}{\zeta^j}\hat{f}=\oint_0\frac{d\zeta}{\zeta^{j-2}}\chi_1
-\oint_\infty\frac{d\zeta}{\zeta^{j}}\chi_2.
\end{equation}
Given $\hat{f}$ we define $G$ such that $\partial G/\partial z(\zeta)=\hat{f}/\zeta^2.$
Next, we define the function of the coefficients of $z(\zeta)$
\begin{equation}
F(z,v,w,\bar{v}, \bar{z})=\frac{1}{2\pi i}\oint_C\frac{d\zeta}{\zeta^2}G.
\end{equation}
The generalized Legendre transform construction generates a formula for a  K\"ahler potential $K(z,\bar{z}, u,\bar{u})$ of the coordinates $z$ and $u$.
We impose the differential constraint $\partial F/\partial w=0$ which determines $w$
as a function of $z$ and $v$, and perform a Legendre transform on $F$  with respect to coordinates $v$ and $\bar{v}$:
\begin{equation}
K(z,\bar{z}, u, \bar{u})=F-v u-\bar{v}\bar{u},\ u=\frac{\partial F}{\partial v},\ \bar{u}=\frac{\partial
F}{\partial \bar{v}}.
\end{equation}

In our case  the above procedure produces
\begin{equation}
\oint_C\frac{d\zeta}{\zeta^j}\hat{f}=-4\mu\oint_0\frac{d\zeta}{\zeta^j}{\zeta}+\left(\oint_0-\oint_\infty\right)\frac{d\zeta}{\zeta^{j-2}}\chi_1,
\end{equation}
since we chose $u_{ij}$ so that (\ref{trivial}) holds, we have
\begin{eqnarray}
&&F(z,v,w,\bar{v}, \bar{z})=-4 \mu \frac{1}{2\pi i}\oint_0\frac{d\zeta}{\zeta}\frac{z(\zeta)}{\zeta^2}\nonumber\\
&&-\frac{1}{2\pi i}\sum_{i=1}^{k}\sum_{l=0,1}\oint_{C^l_i}\frac{d\zeta}{\zeta}\frac{\sqrt{z(\zeta)}+(-1)^lq_i(\zeta)}{\zeta}\log(\sqrt{z(\zeta)}+(-1)^lq_i(\zeta)).\nonumber
\end{eqnarray}
For each $i$ the pair $(C^0_i, C^1_i)$ consists of 4 figure-eight shaped contours on the
Riemann surface $\eta^2=z(\zeta),$ each contour surrounding two out of the eight points
$(\eta,\zeta)=(\pm q_i(\alpha_{ij}),\alpha_{ij}).$ The projection of $C^0_i$ on the
$\zeta$ plane coincides with the projection of $C^1_i.$ The two contours $C^1_i$  have
figure-eight shapes encircling points $\alpha_{ij}$ clockwise for odd $j$ and
counterclockwise for even $j.$
With this in mind we obtain
\begin{equation}\label{intrel}
\sum_{l=0,1}\frac{1}{2\pi i}\oint_{C_i^l}\frac{d\zeta}{\sqrt{z}}f(\zeta)\log(\sqrt{z}+(-1)^l q_i)=
2\sum_{j=1}^4(-1)^j\int^{\alpha_{ij}}\frac{d\zeta}{\sqrt{z}}f(\zeta).
\end{equation}

For convenience let us define matrices
\begin{equation}
\F=\left(\begin{array}{ccc} F_{vv}&F_{vw}&F_{v\bar{v}}\\ F_{wv}&F_{ww}&F_{w\bar{v}}\\ F_{\bar{v}v}&F_{\bar{v}w}&F_{\bar{v}\bar{v}}\end{array}\right),
\end{equation}
where the subscripts of $F$ denote partial derivatives (e.g. $F_{vw}=\partial^2 F/\partial v\partial w$), and
\begin{equation}
\G=\left(\begin{array}{ccc} G^{vv}&G^{vw}&G^{v\bar{v}}\\ G^{wv}&G^{ww}&G^{w\bar{v}}\\
G^{\bar{v}v}&G^{\bar{v}w}&G^{\bar{v}\bar{v}}\end{array}\right)=\F^{-1},
\end{equation}
where the superscripts merely label the components of $\G.$
The components of the metric are
\begin{eqnarray}
K_{z\bar{z}}&=&F_{z\bar{z}}-F_{z a}G^{a b}F_{b\bar{z}}\\
K_{u\bar{z}}&=&F_{\bar{z}a}G^{a v},\ K_{u\bar{u}}=-G^{v\bar{v}}
\end{eqnarray}
where $a$ and $b$ are summed over the values $(v,w,\bar{v}).$ And the metric is
\begin{eqnarray}
ds^2&=&\frac{-1}{G^{v\bar{v}}}\left(-G^{v\bar{v}}(F_{z\bar{z}}-F_{z a'}G^{a' b'}F_{b'\bar{z}}+F_{z a'}G^{a'\bar{v}} (G^{v\bar{v}})^{-1}G^{v b'}F_{b'\bar{z}})dzd\bar{z}+\right.\nonumber\\
&&+\left. (F_{z b} G^{b\bar{v}} dz-G^{v\bar{v}}du)(F_{\bar{z}a}G^{a v}
d\bar{z}-G^{v\bar{v}}d\bar{u})\right)
\end{eqnarray}
Since for $a=v$ as well as for $b=\bar{v}$ we have  $G^{v\bar{v}}G^{ab}=G^{a\bar{v}}G^{v
b}$, we introduce also an index $a'$ taking values $w,\bar{v}$ and $b'$ with values
$v,w.$ The unwieldy  coefficient of $dzd\bar{z}$ simplifies due to the identity (see
Appendix for the proof)
\begin{equation}
\det K_{(z,u)}=\det\left(\begin{array}{cc} F_{z\bar{z}}-F_{z a'}G^{a' b'} F_{b'\bar{z}}& F_{z a'}G^{a'\bar{v}}\\
G^{v b'}F_{b'\bar{z}}& -G^{v\bar{v}} \end{array}\right)=1,
\end{equation}
and the metric has the following simple form
\begin{equation}\label{metric2}
ds^2=\frac{1}{\beta}\left(dzd\bar{z}+(\alpha dz+\beta du)(\bar{\alpha}d\bar{z}+\bar{\beta}d\bar{u})\right)
\end{equation}
where
\begin{equation}
\alpha=F_{z b}G^{b\bar{v}} ,\ \ \beta=-G^{v\bar{v}}.
\end{equation}

To find expressions for the exact form of $\alpha, \beta,$ and $u$ note that
\begin{eqnarray}
u=F_v&=&-\frac{1}{2\pi i}\sum_{i=1}^k\oint_{C^1_i}\frac{d\zeta}{\zeta}\frac{1}{\sqrt{z}}\log(\sqrt{z}-q_i),\\
\bar{u}=F_{\bar{v}}&=&\frac{1}{2\pi i}\sum_{i=1}^k\oint_{C^1_i}\frac{d\zeta}{\zeta}\frac{\zeta^2}{\sqrt{z}}\log(\sqrt{z}-q_i),\\
F_w&=&-4\mu-\frac{1}{2\pi i}\sum_{i=1}^k\oint_{C^1_i}\frac{d\zeta}{\sqrt{z}}\log(\sqrt{z}-q_i).
\end{eqnarray}
If we introduce $p_n$ defined by
\begin{equation}\label{p}
p_n=\frac{1}{2\pi i}\sum_{i=1}^k\oint_{C^1_i}\frac{d\zeta}{2}\frac{\zeta^{n+2}}{z}\left(-\frac{\log(\sqrt{z}-q_i)}{\sqrt{z}}+\frac{1}{\sqrt{z}-q_i}\right)
\end{equation}
then
\begin{equation}
\G^{-1}=\F=\left(\begin{array}{ccc} -p_{-2}&-p_{-1}&p_0\\ -p_{-1}&-p_0&p_1\\ p_0&p_1&-p_2\end{array}\right),
\end{equation}
and
\begin{equation}
F_{zv}=-p_{-3},\ F_{zw}=-p_{-2},\ F_{z\bar{v}}=p_{-1}.
\end{equation}
Let $D=\det F,$ then
\begin{equation}
\alpha=\frac{1}{D}\left|\begin{array}{ccc}
p_{-3}&p_{-2}&p_{-1}\\
p_{-2}&p_{-1}&p_0\\
p_{-1}&p_0&p_1
\end{array}\right|,\ \beta=\frac{1}{D}\left|\begin{array}{cc} p_{-1}&p_0\\ p_0&p_1\end{array}\right|.
\end{equation}

\section{Comparison of the two approaches}

From the form of the one-forms $A_0$ and $B_0$ of equation (\ref{diffs}) combined with the
expression (\ref{abel}, \ref{chi1}) for $\chi$ we can identify $u=\chi(0)$ and
\begin{equation}
\overline{A_0}=(2\mu-l)\frac{d z}{z},\ \overline{B_0}=d u-(L(0)-u)\frac{d z}{z}.
\end{equation}
The two expressions (\ref{metric1}),(\ref{metric2}) for the metric coincide if
\begin{eqnarray}\label{abgd}
\alpha&=&\left(\gamma-(L(0)-u)\frac{\delta}{2\mu-l}\right)e^{i\phi},\\
\beta&=&z\frac{\delta}{2\mu-l} e^{i\phi}
\end{eqnarray}
for some real-valued function $\phi$.
 To start, we use the expressions (\ref{gammadelta}) for $\delta$ and $\gamma$ to find
\begin{eqnarray}
\frac{\delta}{2\mu-l}&=&\frac{[1,a,a C,a^2C]}{[(2\mu-l)a+a^2C,1,a C,C]}\\
\gamma-\frac{(L(0)+u)\delta}{2\mu-l}&=&-\frac{[L(0)+u-C,a,a C,a^2C]}{[(2\mu-l)a+a^2C,1,a C,C]}
\end{eqnarray}

On the other hand, introducing $\Pi_l=\prod_{j\neq l}(a_l-a_j)$  and using  (\ref{p}), one finds
\begin{eqnarray}
p_2&=&-\frac{2\mu-l}{\bar{z}}-\frac{1}{\bar{z}}\sum_{l=1}^4 \frac{a_l^{4}}{\Pi_l}C_l,\\
p_n&=&-\frac{1}{\bar{z}}\sum_{l=1}^4 \frac{a_l^{n+2}}{\Pi_l}C_l,\ -2\leq n\leq1\\
p_{-3}&=&-\frac{1}{\bar{z}}\frac{1}{a_1a_2a_3a_4}(L(0)-\frac{1}{2}\chi(0))-\frac{1}{\bar{z}}\sum_{l=1}^4 \frac{1}{a_l\Pi_l}C_l,
\end{eqnarray}
which leads to the following expressions for the determinants (see Appendix)
\begin{equation}
\bar{z}^3\left|\begin{array}{ccc}
p_{-3}&p_{-2}&p_{-1}\\
p_{-2}&p_{-1}&p_0\\
p_{-1}&p_0&p_1
\end{array}\right|=\frac{1}{a_1a_2a_3a_4}\frac{[L(0)-\frac{1}{2}\chi(0)-C,a,aC,a^2C]}{[1,a,a^2,a^3]},
\end{equation}
\begin{equation}
p_{-1}p_1-p_0^2=\frac{-1}{\bar{z}^2}\frac{[1,a,aC,a^2C]}{[1,a,a^2,a^3]},
\end{equation}
\begin{equation}
\bar{z}^3 D=-\frac{[(2\mu-l)a+a^2C,1,aC,C]}{[1,a,a^2,a^3]}.
\end{equation}
Indeed, recalling that $a_1a_2a_3a_4=z/\bar{z}$ we find
\begin{eqnarray}
\alpha&=&-\frac{\bar{z}}{z}\frac{[L(0)-\frac{1}{2}\chi(0)-C,a,a C,a^2C]}{[(2\mu-l)a+a^2C,1,a C,C]},\\
\beta&=&\bar{z}\frac{[1,a,a C,a^2C]}{[(2\mu-l)a+a^2C,1,a C,C]}.
\end{eqnarray}
Thus the relations (\ref{abgd}) indeed hold with $e^{i\phi}=\bar{z}/z.$

\section*{Acknowledgments}
SCh was supported in part by DOE grant DE-FG02-90ER40542. NJH thanks the University of Bonn and SFB40 (now the Max-Planck-Institut) for its support 24 years ago.

\section*{Appendix}
\subsection*{Resultants}
Here we find the condition on coefficients $\C_i$ and $\D_i$ for
the following system of equations to have a solution in $\zeta$
$$\sum_{j=1}^4\frac{\C_j}{\zeta-a_j}=0,\quad
\sum_{j=1}^4\frac{\D_j}{\zeta-a_j}=0.
$$

To compare with the expressions of Section \ref{canonical} put $\C_j=C_j da_j$ and
$\D_j=D_j da_j.$ For convenience let us introduce polynomials
$h(\zeta)=\prod_{i=1}^4(\zeta-a_i)$ and $g_j(\zeta)=h(\zeta)/(\zeta-a_j).$ Then for
\begin{equation}
M(\zeta)=\sum_{i=1}^4 \C_i g_i(\zeta),\ N(\zeta)=\sum_{i=1}^4 \D_i g_i(\zeta),
\end{equation}
the above system is equivalent to the system of two third order equations
$M(\zeta)=0=N(\zeta).$ $M$ and $N$ have no common root if and only if the six polynomials
$M(\zeta), \zeta M(\zeta), \zeta^2 M(\zeta), N(\zeta), \zeta N(\zeta), \zeta^2 N(\zeta)$
are linearly independent. These are fifth order polynomials and can be expanded in the
basis formed by e.g. $h(\zeta), \zeta h(\zeta), g_i(\zeta).$

It is convenient to introduce
\begin{equation}
A_0=\sum_{i=1}^4 \C_i,\ B_0=\sum_{i=1}^4 \D_i,\ A_1=\sum_{i=1}^4
a_i \C_i,\ B_1=\sum_{i=1}^4 a_i \D_i,
\end{equation}
then
\begin{eqnarray} \zeta M(\zeta)&=&A_0 h(\zeta)+\sum a_j
\C_j g_j(\zeta)\\
\zeta^2 M(\zeta)&=&A_0 \zeta h(\zeta)+A_1 h(\zeta)+\sum a_j^2 \C_j g_j(\zeta),
\end{eqnarray}
with analogous expressions for $N(\zeta).$

Now the condition for the existence of a solution is the
vanishing of the Sylvester determinant
\begin{equation}
R=\left|\begin{array}{cccccc}
B_0 & B_1 & a_1^2\D_1 &
a_2^2\D_2 &a_3^2\D_3 &a_4^2\D_4 \\
0 & B_0 & a_1 \D_1 & a_2 \D_2 & a_3 \D_3 & a_4
\D_4\\
0 & 0 & \D_1 & \D_2 & \D_3 & \D_4\\
A_0 & A_1 & a_1^2\C_1 &
a_2^2\C_2 &a_3^2\C_3 &a_4^2\C_4 \\
0 & A_0 & a_1 \C_1 & a_2 \C_2 & a_3 \C_3 & a_4
C_4\\
0 & 0 & \C_1 & \C_2 & \C_3 & \C_4\\
\end{array}\right|
\end{equation}
One finds $R$ to be given by
\begin{equation}
R=\ep^{ijkl}\C_i\D_ja_k(A_0\D_k-B_0\C_k)\left[a_l^2(A_0\D_l-B_0\C_l)-a_l(B_1\C_l-A_1\D_l)\right].
\end{equation}
Introducing $A_2=\sum_{i=1}^4 \C_i/a_i$ and $B_2=\sum_{i=1}^4 \D_i/a_i$ we have
\begin{equation}
R=\left|\begin{array}{cccccc}
B_0 & 0 & a_1^2\D_1 &
a_2^2\D_2 &a_3^2\D_3 &a_4^2\D_4 \\
0 & 0 & a_1 \D_1 & a_2 \D_2 & a_3 \D_3 & a_4
\D_4\\
0 & -B_2 & \D_1 & \D_2 & \D_3 & \D_4\\
A_0 & 0 & a_1^2\C_1 &
a_2^2\C_2 &a_3^2\C_3 &a_4^2\C_4 \\
0 & 0 & a_1 \C_1 & a_2 \C_2 & a_3 \C_3 & a_4
\C_4\\
0 & -A_2 & \C_1 & \C_2 & \C_3 & \C_4\\
\end{array}\right|
\end{equation}

\subsection*{Determinant relations}
Computing the determinant of the $8\times8$ matrix
\begin{equation}
\left|\begin{array}{cc} aC&0\\ C&C/a\\ a^2D&aD\\aD&0\\0&D\\D&D/a\\0&C\\0&aC\end{array}\right|=
\left|\begin{array}{cc} aC&0\\ 0&C/a\\ 0&aD\\aD&0\\-aD&D\\0&D/a\\-aC&C\\-a^2C&aC\end{array}\right|=
\left|\begin{array}{cc} aC&0\\ 0&C/a\\ 0&aD\\aD&0\\0&D\\0&D/a\\0&C\\-a^2C&aC\end{array}\right|=0.
\end{equation}
On the other hand the same determinant equals
\begin{eqnarray}
&&-[a C,C,a^2D,D][a C,C,D,D/a]+[C/a,C,a C,D][a C,a^2D,a D,D]\nonumber\\
&&+[a C,C,a D,D]^2.
\end{eqnarray}
Thus
$$
[a C,C,a^2D,a D]\overline{[a^2C,a C,a D,D]}=$$
\begin{equation}
[C,a C,a^2C,a D]\overline{[C,a C,a^2C,a D]}+[C,a C,D,a D]^2
\label{determinants}
\end{equation}
which produces equation (\ref{relation}) used in Section \ref{canonical}.

\subsection*{The determinant of $ K_{(z,u)}$ }
The function $z(\zeta)$ trivially satisfies the relations
$z(\zeta)_{z\bar{z}}=-z(\zeta)_{v\bar{v}},z(\zeta)_{z\ap}=z(\zeta)_{v(\ap-1)},
z(\zeta)_{\bp\bz}=-z(\zeta)_{(\bp+1)\bv}$, which implies analogous relations for $F.$
\begin{equation}
\det
K_{(z,u)}=-G^{v\bv}F_{z\bz}+F_{z\ap}\left(G^{v\bv}G^{\ap\bp}-G^{\ap\bv}G^{v\bp}\right)F_{\bp\bz}.
\end{equation}
For an $n\times n$ matrix $A$ we denote by $A^{(k)}$ the $k$-th compound matrix, which is
the matrix composed of all order $k$ minors of $A$. The adjugate compound matrix
$\adj^{(k)}$ is obtained from $A^{(k)}$ by replacing each $k$-minor by its complementary
minor with the corresponding factor of $(-1)^l$ and transposition. In other words, due to
the Laplace expansion of determinants, $A^{(k)}\adj^{(k)}A=\det A\ {\bf 1}.$ Thus from
the Binet-Cauchy theorem
\begin{equation}
G^{(2)}=\left(F^{(2)}\right)^{-1}=\frac{1}{\det F}\adj^{(2)} F.
\end{equation}
Thus utilizing the relations $F_{z\bz}=-F_{v\bv}, F_{z\ap}=F_{v(\ap-1)},$ and $F_{\bp\bz}=-F_{(\bp+1)\bv}$ we have
\begin{equation}
\det K_{(z,u)}=\frac{1}{\det F}\left(F_{v\bv} F^{(1)}_{v\bv}-F_{v(\ap-1)}F_{(\bp+1)\bv}\adj^{(2)}F_{(v,\ap;\bp,\bv)}\right).
\end{equation}
The expression in brackets above is exactly the Cauchy expansion for $\det F.$ Thus we have $\det K_{(z,u)}=1.$

\subsection*{Relations between the $p_n$ and $C_l$ expressions}
We observe that
\begin{equation}
[1,a,a^2,a^3]=\epsilon^{ijkl}\Pi_i\Pi_j\frac{a_k-a_j}{a_j-a_i},
\end{equation}
without summation over the repeated indices.
Then
\begin{eqnarray}
&&(2\bar{z})^2(p_{-1}p_1-p_0^2)=\sum_{i,l}\frac{a_l-a_i}{\Pi_i\Pi_l}a_iC_ia_l^2C_l=\nonumber\\
&&=\frac{-1}{2}\frac{1}{[1,a,a^2,a^3]}\epsilon^{ijkl}(a_k-a_j)a_iC_ia_l^2C_l=
-\frac{[1,a,a C,a^2C]}{[1,a,a^2,a^3]}.
\end{eqnarray}

Next note the identities
\begin{equation}
\sum_{l=1}^4 \frac{a_l^k}{\Pi_l}=\left[\begin{array}{l}
1,\ k=3\\
0,\ 0\leq k\leq 2\\
-{1}/{a_1a_2a_3a_4},\ k=-1\end{array}\right.
\end{equation}
Using these we have

$$-(2\bar{z})^3\left|\begin{array}{ccc}
p_{-3}&p_{-2}&p_{-1}\\
p_{-2}&p_{-1}&p_0\\
p_{-1}&p_0&p_1
\end{array}\right|=$$
$$
=\det\left(\begin{array}{c} 1/a\\1\\a\\a^2\end{array}\right)
\diag(\frac{1}{\Pi_1},\frac{1}{\Pi_2},\frac{1}{\Pi_3},\frac{1}{\Pi_4})
(C-L(0)+\frac{1}{2}\chi(0), a C, a^2C, a)=
$$
$$
=-\frac{1}{a_1a_2a_3a_4}\frac{[L(0)-\frac{1}{2}\chi(0)-C,a,a C,a^2C]}{[1,a,a^2,a^3]},$$

as well as
\begin{eqnarray}
(2\bar{z})^3 D&=&\det \left(\begin{array}{c} 1\\a\\a^2\\a^3\end{array}\right)
\diag(\frac{1}{\Pi_1},\frac{1}{\Pi_2},\frac{1}{\Pi_3},\frac{1}{\Pi_4})
\left(C,a C, a^2C+a(2\mu-l), 1\right)=\nonumber\\
&=&-\frac{[(2\mu-l)a+a^2C,1,a C, C]}{[1,a,a^2,a^3]}.
\end{eqnarray}

\end{document}